\documentclass[a4paper]{IEEEtran}

\ifCLASSINFOpdf
\else
\fi
\usepackage{amsmath,amssymb,bm}
\usepackage{mathtools,nccmath}
\usepackage{algorithmic}

\usepackage{array}
\usepackage{tikz}
\usepackage{tikzscale}
\usepackage{pgf}
\usepackage[T1]{fontenc}
\usepackage{mathtools}
\usepackage[numbers,sort&compress]{natbib}
\usepackage[linesnumbered,lined,ruled,commentsnumbered]{algorithm2e}
\usepackage{lipsum}  
\usepackage{authblk}

\IEEEoverridecommandlockouts
\begin{document}
%
\title{Sum-rate Maximization in Uplink CRAN with a Massive MIMO Fronthaul}
%
%
%
\author[$\ast,\dagger$]{Dick~Maryopi }
\author[$\S$]{Yingjia~Huang}
\author[$\S$]{Aissa~Ikhlef
\thanks{This work was supported by the UK Engineering and Physical Sciences Research Council under Grant EP/R044090/1.}}
\affil[$\ast$]{School of Electrical Engineering, Telkom University,  Bandung, Indonesia, maryopi@telkomuniversity.ac.id\vspace{.5ex}}
\affil[$\S$]{Department of Engineering, Durham University, Durham, UK, \{yingjia.huang, aissa.ikhlef\}@durham.ac.uk \vspace{.5ex}}
\affil[$\dagger$]{The University Center of Excellence for Advanced Intelligent Communications (AICOMS), \protect\\ Telkom University,  Bandung, Indonesia.}


\markboth{Preprint}%
{Shell \MakeLowercase{\operatornameit{et al.}}: Bare Demo of IEEEtran.cls for IEEE Journals}
%



\maketitle

\begin{abstract}
The limited fronthaul capacity is known to be one
of the main problems in cloud radio access networks (CRANs),
especially in the wireless fronthaul links. In this paper, we
consider the uplink of a CRAN system, where massive multiple-input
multiple-output (MIMO) is utilized in the fronthaul link.
Considering multi-antenna user equipment (UEs) and multi-antenna
remote radio heads (RRHs), we maximize the system
sum-rate by jointly optimizing the precoders at the UEs and the
quantization noise covariance matrices and transmit powers at
the RRHs. To solve the resulting nonconvex problem, an iterative
algorithm based on the majorization-minimization (MM) method
is proposed. Two schemes at the central unit are considered,
namely maximum ratio (MR) and zero-forcing (ZF) combining.
Numerical results show that the sum-rate has an asymptotic
behaviour with respect to the maximum available power at RRHs
and that the MR scheme goes to its asymptote faster than the
ZF scheme.


\end{abstract}


%
\IEEEpeerreviewmaketitle

\section{Introduction}
%

In future wireless networks, delivering a high data rate will remain the main driver along with other features such as high reliability, low latency and energy efficiency while supporting a large number of users in a scalable fashion. Among the potential enablers for that is the cloud radio access network (CRAN) architecture, which also has been considered in the fifth generation (5G) and beyond of wireless networks. The primary idea of CRAN is to pool the baseband unit (BBU) of several base stations altogether in a central unit (CU) rather than place them separately in each base station \cite{quek_peng_simeone_yu_2017}. By migrating the baseband processing from the base stations to the CU, low-complexity and low-cost base stations, also called remote radio heads (RRHs), can then be managed to cooperate to effectively suppress interference resulting in higher spectral and energy efficiencies. To exploit this, the RRHs should be distributed over the service area and connected to the CU through fronthaul links.

In the CRAN architecture, the fronthaul links play a crucial role because they constrain the amount of data that can be transmitted from the user equipment (UE) to the CU and vice versa. Since it is a baseband signal with a very high load that should be transmitted over the fronthaul links, they can become a bottleneck even if a fibre optic cable is deployed \cite{6924850}. In this regard, many works have investigated the impact of limited-capacity fronthaul in CRAN and proposed compression and optimization schemes to resolve the problem \cite{quek_peng_simeone_yu_2017, 6924850, 6588350, 7590010, 7573000, 7797237}. However, most of the works considered the wired fronthaul links, which could be unsuitable for certain deployment scenarios. For example, in difficult terrains and rural areas, the use of the wireless medium for fronthaul links might be preferable due to its lower deployment cost and its flexible deployment and adjustment. Together with the emergence of millimeter wave technology and unmanned aerial vehicle (UAV) techniques, wireless fronthaul links for CRAN is gaining momentum \cite{8660693, 9497357, 9165882}.
 
In this work, we consider a CRAN architecture that utilizes massive MIMO for the wireless fronthaul links. While massive MIMO has been widely studied and is well established for the access links in the cellular system, to the best of authors' knowledge, only \cite{9497357} studied massive MIMO for the fronthaul links in the context of the downlink of a CRAN system with UAVs deployed as flying RRHs. Other than this work, massive MIMO fronthaul has only been studied for small-cell networks \cite{SIG-093, 7070656}, which can differ from CRAN in the baseband processing and user association. Therefore, in those works, compression schemes for dealing with the limited capacity fronthaul is not considered in contrast to our work. In this regard, our motivation for utilizing massive MIMO is straightforward, namely, to provide higher fronthaul capacity while enabling spatial multiplexing to establish the fronthaul links to all RRHs simultaneously.

To utilize massive MIMO fronthaul in the uplink, we involve all CRAN entities from the UEs to the CU in our system design. In particular, we aim to jointly optimize the precoding matrices at the UEs, the quantization covariance matrices and transmit powers at the RRHs to maximize the weighted sum-rate. Thereby, the problem is formulated as a joint optimization problem for the fronthaul and access links under transmit power constraints at the UEs and RRHs. Unlike most of the existing works in CRAN, where the wired fronthaul is assumed as an ideal bit-pipe with a fixed capacity constraint, we have in this paper a varying capacity constraint due to the dynamic of non-negligible impairments that naturally exists in a wireless channel such as interference and fading. This constraint depends on the schemes applied at the far end of the fronthaul link at the CU, where we investigate in this work both the maximum ratio (MR) and the zero-forcing (ZF) combining schemes for the sake of their low complexity.
However, the formulated problem is challenging to solve due to the nonconvexity in the objective function and constraints. To handle this problem, we propose an algorithm based on the majorization-minimization (MM) approach \cite{beck_teboulle_2009}. For each investigated scheme, MR and ZF, the algorithm is shown numerically to converge.  

\begin{figure}[!t]
\centering
\usetikzlibrary{arrows}
\usetikzlibrary{positioning}
\usetikzlibrary{shadows}
\usetikzlibrary{shapes.geometric}

\tikzstyle{block1} = [draw, fill=blue!20, rectangle, text centered, 
     minimum height=2em, text width=3em, font=\sffamily]
 \tikzstyle{block2} = [draw, fill=blue!20, circle, 
    minimum height=.25em, minimum width=.2em]
 \tikzstyle{block3} = [draw, fill=blue!20, rectangle, 
    minimum height=2em, minimum width=2em]
\tikzstyle{input} = [coordinate]
\tikzstyle{output} = [coordinate]
\tikzstyle{pinstyle} = [pin edge={to-,thin,black}]
\tikzstyle{block4}=[draw, minimum height=1.5em, minimum width=2em, font=\sffamily]

\tikzset{naming/.style={align=center,font=\small}}
\tikzset{antenna/.style={insert path={-- coordinate (ant#1) ++(0,0.25) -- +(135:0.25) + (0,0) -- +(45:0.25)}}}
\tikzset{station/.style={naming,draw,shape=dart,shape border rotate=90, minimum width=10mm, minimum height=10mm,outer sep=0pt,inner sep=3pt}}
\tikzset{mobile/.style={naming,draw,shape=rectangle,minimum width=15mm,minimum height=7.5mm, outer sep=0pt,inner sep=3pt}}
\tikzset{mobile/.style={naming,draw,shape=rectangle,minimum width=12mm,minimum height=6mm, outer sep=0pt,inner sep=3pt}}
\tikzset{radiation/.style={{decorate,decoration={expanding waves,angle=90,segment length=4pt}}}}

\tikzset{%
  wireless/.pic={
      \draw [-<, thick] (0,0) -| (.5,#1);
    \foreach \r in {.1,.2,.3}
     \draw (.6,#1) ++ (60:\r) arc (60:-60:\r);
  },
  vdots/.pic={
    \foreach \i in {-.25,-.125, 0}
      \fill (.25,\i) circle [radius=.75pt]; 
  },
  block/.style={
    shape=rectangle,  rounded corners=5pt,
    minimum width=1.25cm,
    minimum height=1.25cm,
    draw, thick,font=\sffamily,
    draw=gray, thick},
  Rx/.style 2 args={
    block,
    node contents={},
    append after command={
      \pgfextra{\pgfnodealias{@}{\tikzlastnode}}
      (@.north #1) [yshift=-.25cm] pic [#2] {wireless=.5}
   	(@.#1)                        pic [#2] {vdots}
   	(@.south #1) [yshift= .25cm] pic [#2] {wireless=.5}
    }
  },
  MIMO Rx west/.style={Rx={west}{xscale=-1}},
}

\tikzset{%
  wireless/.pic={
      \draw [-<, thick] (0,0) -| (.5,#1);
    \foreach \r in {.1,.2,.3}
     \draw (.6,#1) ++ (60:\r) arc (60:-60:\r);
  },
  vdots/.pic={
    \foreach \i in {-.25,-.125, 0}
      \fill (.25,\i) circle [radius=.75pt]; 
  },
  block/.style={
    shape=rectangle, rounded corners=5pt,
    minimum width=1.25cm,
    minimum height=1.55cm,
    draw, thick,font=\sffamily,
   fill=blue!15, thick},
  Tx/.style 2 args={
    block,
    node contents={},
    append after command={
      \pgfextra{\pgfnodealias{@}{\tikzlastnode}}
      (@.north #1) [yshift=-.75cm] pic [#2] {wireless=.5}   	
    }
  },
  MIMO Tx east/.style={Tx={east}{xscale=1}},
}

\tikzstyle{output} = [coordinate]
\tikzstyle{pinstyle} = [pin edge={to-,thin,black}]


\newcommand{\MBS}[1]{%
\begin{tikzpicture}[thick]
\node[station] (base) {#1};

\draw[line join=bevel] (base.100) -- (base.80) -- (base.110) -- (base.70) -- (base.north west) -- (base.north east);
\draw[line join=bevel] (base.100) -- (base.70) (base.110) -- (base.north east);

 original yshift=0.8pt
\draw[line cap=rect] ([xshift=.5cm,yshift=.3pt] base.north) [antenna=1];
\draw[line cap=rect] ([yshift=.3pt]ant1 |- base.north) -- node[above,shape=rectangle,inner ysep=+.3333em]{\dots} ([xshift=-.5cm,yshift=.3pt]base.north) [antenna=2];
\end{tikzpicture}
}

\newcommand{\MUE}[1]{%
\begin{tikzpicture}[every node/.append style={rectangle,minimum width=0pt}, thick]
\node [mobile,label={[inner ysep=+.3333em]\dots}, font=\small\sffamily, thick, rounded corners=5pt] (box) {#1};

\draw ([xshift=.25cm] box.north west) [antenna=1];
\draw ([xshift=-.25cm]box.north east) [antenna=2];
\end{tikzpicture}
}

\newcommand{\RRH}[1]{%
\begin{tikzpicture}[every node/.append style={rectangle,minimum width=1pt}, thick]
\node [MIMO Rx west];
\node [MIMO Tx east];
\end{tikzpicture}
}

\resizebox{\columnwidth}{!}{%
\begin{tikzpicture}[scale=0.45, auto, node distance=2.85cm, very thick,  draw]
 \node [name=input0] {\MUE{UE$_1$}};       
 \node [name=input1, below of = input0,  node distance = 2.5cm]{\MUE{UE$_k$}};  
\node [name=input2, below of = input1, node distance = 2.55cm] {\MUE{UE$_K$}};   
\node [pinstyle, name=dot1, below of = input0, node distance = 1.1cm] {$\vdots$};   
\node [pinstyle, name=dot2, below of = input1, node distance = 1.1cm] {$\vdots$};  

\node [ right of=input0, node distance = 6cm] (g0) {\RRH{}};
\node at (g0)[ node distance = 6cm,font=\sffamily] {RRH$_1$};
\node [ right of=input1, node distance = 6cm] (g1) {\RRH{}};
\node at (g1)[ node distance = 6cm,font=\sffamily] {RRH$_l$};
\node [ right of=input2, node distance = 6cm] (g2) {\RRH{}};
\node at (g2)[  node distance = 6cm,font=\sffamily]  {RRH$_L$};
 \node [pinstyle, name=dot3, below of = g1, node distance = 1.275cm] {$\vdots$};   
 \node [pinstyle, name=dot4, below of = g0, node distance = 1.275cm] {$\vdots$};   
  
 \node [input, name=outputg0, right of= g0, node distance = 1.8cm] {};       
 \node [input, name=outputg1, right of = g1, node distance = 1.8cm] {};  
\node [input, name=outputg2, right of = g2, node distance = 1.8cm] {};   
\node[ right of =outputg1, node distance = 3cm, font=\sffamily](plus){\MBS{CU}};

\node[pinstyle, name = annotation1, below of = input2, xshift=-.5cm, node distance = 1.75cm]{}; 
\node[pinstyle, name = annotation2, right of = annotation1, node distance = 2.5cm]{ : Access link}; 
\node[pinstyle, name = annotation3, right of = annotation2, node distance = 2.5cm]{}; 
\node[pinstyle, name = annotation4, right of = annotation3, node distance = 2.5cm]{ : Fronthaul link}; 

\path[->, dashed, color=gray!]
   (input0)edge(g0)
   (input0)edge(g1)
   (input0)edge(g2)     
  ; 
  \path[->, dashed, color=gray!]
   (input1)edge(g0)
   (input1)edge[color=black]  node[above left, color=black!]{$\mathbf{H}_{lk}$}(g1)
   (input1)edge(g2)       
  ; 

\path[->, dashed, color=gray!]
   (input2)edge(g0)
   (input2)edge(g1)
   (input2)edge(g2)
  ; 

  \path[->,thick]     
	(outputg0)edge node[below=.15cm ]{$\mathbf{g}_1$} (plus) 
	(outputg1)edge node[below ]{$\mathbf{g}_l$} (plus)  
	(outputg2)edgenode[below =.2cm]{$\mathbf{g}_L$}(plus) 
 ;
 
 \path[->, dashed, color=gray!]
   (annotation1)edge(annotation2)   
  ;   
  
 \path[->,thick]
   (annotation3)edge(annotation4)   
  ; 
   
\end{tikzpicture}
}
\caption{The schematic diagram of the C-RAN system utilizing massive MIMO fronthaul.}
\label{SystemModel}
\end{figure}
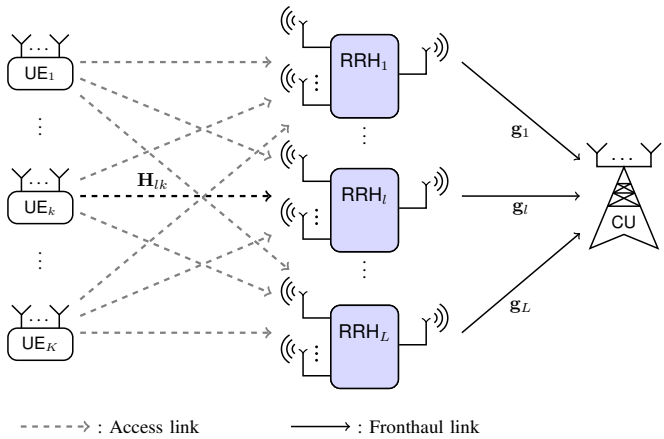

The rest of this paper is organized as follows. In Section \ref{sysMod}, we describe the system model. In Section \ref{cranDesign}, we present our proposed CRAN design, in which we state the optimization problem and provide the proposed solution. Numerical results are provided in Section \ref{numResults}. We conclude the paper in Section \ref{conclusion}.

\section{System Model}\label{sysMod}

We consider the uplink of a CRAN system, where $K$ multi-antenna UEs wish to wirelessly send their messages to the CU in two hops through $L$ multi-antenna RRHs, as depicted in Fig. \ref{SystemModel}. The CU is equipped with a massive uniform linear array (ULA) of size $N_C$. For $k\in \mathcal{K}=\{1, \dots, K\}$, the $k$-th user is equipped with $N_{U,k}$ antennas, which brings on the total number of antennas at all UEs to $N_U=\sum_{\mathcal{K}} N_{U,k}$. For $l\in \mathcal{L}=\{1, \dots, L\}$, the $l$-th RRH utilizes $N_{H,l}$ antennas for reception, with the total number of antennas at all RRHs for reception is  $N_H=\sum_{\mathcal{L}}N_{H,l}$, and one antenna for transmission. This is because in the fronthaul links we have a massive MIMO channel and hence the assumption of using single antenna RRHs for transmission in the fronthaul is of practical interest. We assume that the access links (UEs-RRHs links) and fronthaul links (RRHs-CU links) are separated in the time domain to avoid interference between them. %
\subsection{Access Links}
Let $M_k\in\{1, \cdots, 2^{nR_k}\}$ denote the message to be transmitted by UE $k$ to the CU, where  $n$ is the block length and $R_k$ is the information rate in bits per channel use (bpcu). Then, the $k$-th UE encodes its message using a Gaussian codebook into a data stream $\mathbf{s}_k \in \mathbb{C}^{d_k\times1}$ of unit variance and transmits it using a linear precoder $\mathbf{V}_k\in\mathbb{C}^{N_{U,k}\times d_k}$. The resulting signal $\mathbf{x}_k=\mathbf{V}_k \mathbf{s}_k$, with power constraint $\mathbb{E}\{\|\mathbf{x}_k\| ^2\}\leq P_{UE}$, is then transmitted through the channel $\mathbf{H}_{lk}\in \mathbb{C}^{N_{H,l}\times N_{U,k}}$ to the $l$-th RRH.  A Rayleigh flat-fading channel model is used for $\mathbf{H}_{lk}$, i.e., $\mathbf{H}_{lk} = \sqrt{\tilde{\beta}_{lk}} \mathbf{\tilde{H}}_{lk}$, where $\tilde{\beta}_{lk}$ is the large-scale fading coefficient of the channel between UE $k$ and RRH $l$, and  $\mathbf{\tilde{H}}_{lk}$ is a matrix of independent Rayleigh coefficients with entries modeled as $\mathcal{CN}(0, 1)$. The received signal at the $l$-th RRH from all UEs is then given by
\begin{align}\label{accesLink}
    \mathbf{y}_{H,l}=\sum_{k=1}^{K} \mathbf{H}_{lk}\mathbf{x}_k+\mathbf{z}_{H,l},
\end{align}
where $\mathbf{z}_{H,l}\sim\mathcal{CN}(\mathbf{0, I}_{N_{H}})$ is the additive noise at the $l$-th RRH. Next, the received signal $\mathbf{y}_{H,l}$ is transferred to the CU via a wireless fronthaul link. 
\subsection{Fronthaul Links}
%
Similar to \cite{7465790}, we adopt the compress-and-forward relaying strategy at the RRHs. So, before forwarding to the CU via the fronthaul links, the signal $\mathbf{y}_{H,l}$ at the $l$th RRH is quantized and compressed. We adopt the Gaussian quantization test channel to model the quantization process \cite{6588350}, and hence the resulting quantized signal, $\widehat{\mathbf{y}}_{H,l}$, can be written as
\begin{align}
\widehat{\mathbf{y}}_{H,l}=\mathbf{y}_{H,l}+\mathbf{q}_l, \label{testChannelData}
\end{align}
where $\mathbf{q}_l\sim\mathcal{CN}(\mathbf{0, \Omega}_l)$ is the quantization noise, which is independent of $\mathbf{y}_{H,l}$.  Next, $\widehat{\mathbf{y}}_{H,l}$ is compressed to generate the compression index  $T_{l}\in \{1, \dots, 2^{nC_l}\}$, where $C_l$ is the rate. Here, we assume point-to-point compression. The index $T_{l}$ is then mapped into a complex scalar symbol $y_{C,l}$ to be further transmitted by the $l$-th RRH, using a single antenna, to the CU.

The received signal at the CU can be written as
\begin{align}
\mathbf{\widetilde{y}}_{C}=\sum_{l=1}^L \mathbf{g}_{l}y_{C,l}+\mathbf{z}_{C},\label{jAntennaCU}
\end{align}
where $\mathbf{g}_{l}\in\mathbb{C}^{N_C\times 1}$ is the channel between the $l$-th RRH and the CU, $\mathbf{z}_{C}\sim\mathcal{CN}(0, \sigma^2\mathbf{I}_{N_C})$  is the additive noise at the CU and $\mathbb{E}\{\vert y_{C,l}\vert^2\}= p_{H,l}\leq P_{H}^{\operatorname{max}}$ is the transmit power constraint per RRH. Considering a scenario with line-of-sight (LoS) propagation, the channel vector $\mathbf{g}_l$ can be given by \cite{SIG-093, 7973186}
\begin{align}
\mathbf{g}_l =\sqrt{\beta_l}\, \mathbf{b}_l,
 \label{LoSgl}
\end{align}
where $\beta_l$ is the large scale fading from the $l$-th RRH and $\mathbf{b}_l=\left[1 \quad e^{j2\pi \frac{\Delta}{\lambda} \operatorname{sin}(\theta_l)}\quad \dots\quad  e^{j2\pi (N_C-1)\frac{\Delta}{\lambda}\operatorname{sin}(\theta_l)} \right]^T$. Here, $\Delta$ is the antenna spacing, $\lambda$ is the carrier wavelength and $\theta_l$ is the angle of arrival from the $l$-th RRH distributed uniformly in $[0, 2\pi)$. At the CU, an estimate of the received symbol $\widehat{y}_{C,l}$ is obtained from (\ref{jAntennaCU}) using beamformer $\mathbf{u}_l \in \mathbb{C}^{N_C}$, which is designed based on the perfectly known channel information available at the CU.

We further assume that the user codebooks and the RRH codebooks are available at the CU. Consequently, a lossless decoding from the detected symbol $\widehat{y}_{C,l}$ to an index $T_{l}$ can be realized. 
%
\section{The Proposed C-RAN Design}\label{cranDesign}
%
Under the model given in the previous section, we now discuss our proposed design, where we use an optimization framework. The main question is how to jointly optimize the access and fronthaul links to maximize the system sum-rate. To that end, before we formulate the optimization problem, we first need to specify the fronthaul receive combining scheme and the closed-form expression of the achievable rate per user. Here, we assume that the perfect CSI for the access and fronthaul links is available at the CU.
%
\subsection{Fronthaul Receive Combining}
%
At the CU, the first step is to detect the signals $y_{C,l}$, for all $l\in\mathcal{L}$, sent by the RRHs from the received signal $\mathbf{\widetilde{y}}_{C}$ in (\ref{jAntennaCU}). 
In particular, to detect the signal sent by the $l$th RRH, $y_{C,l}$, a combining vector $\mathbf{u}_l$ is applied to the received signal at the CU and the detected signal is given by
\begin{align}
\widehat{y}_{C,l}&=\mathbf{u}_l^{\dagger}\mathbf{g}_l y_{C,l}+\sum_{l^{\prime}\neq l}\mathbf{u}_l^{\dagger}\mathbf{g}_l^{\prime} y_{C,l^{\prime}}+\mathbf{u}_l^{\dagger}z_{C,l}. \label{GeneralUbeam}
\end{align}
Thus, we can compute the signal-to-interference-and-noise ratio (SINR) of the $l$-th RRH as
\begin{align}\label{generalBeamformer}
\operatorname{SINR}_l=\frac{\vert \mathbf{u}_l^{\dagger}\mathbf{g}_l\vert^2 p_{H,l}}{\sigma^2\Vert \mathbf{u}_l\Vert^2+\sum_{l^{\prime}\neq l}\vert \mathbf{u}_{l^{\prime}}^{\dagger}\mathbf{g}_{l^{\prime}}\vert^2 p_{H,l^{\prime}}}. 
\end{align}
Therefore, the rate $C_l$ between RRH $l$ and the CU is achievable if the condition
\begin{align}\label{beamformingCondition}
\alpha_l( \{p_{H,l}\})&\triangleq\operatorname{log}_2\left(1+\frac{\vert \mathbf{u}_l^{\dagger}\mathbf{g}_l\vert^2 p_{H,l}}{\sigma^2\Vert \mathbf{u}_l\Vert^2+\sum_{l^{\prime}\neq l}\vert \mathbf{u}_{l^{\prime}}^{\dagger}\mathbf{g}_{l^{\prime}}\vert^2 p_{H,l^{\prime}}}\right)\nonumber\\
&\geq C_l.
\end{align}
is satisfied. 

Later in Section \ref{numResults}, we will provide a showcase for two popular and simple combining techniques which are MR and ZF processing. The combining matrix, $\mathbf{U}\triangleq\left[\mathbf{u}_1, \dots, \mathbf{u}_L\right]$, is given by
\begin{align}
\mathbf{U} = \begin{cases}
\mathbf{B}^{\dagger}, & \text{for}~ \text{MR}\\
(\mathbf{B}^{\dagger}\mathbf{B})^{-1}\mathbf{B}^{\dagger}, & \text{for}~ \text{ZF},\\
\end{cases} \label{detector}
\end{align}
with $\mathbf{B}\triangleq\left[\mathbf{b}_1, \dots, \mathbf{b}_L\right]$. Plugging (\ref{detector}) into (\ref{generalBeamformer}), the fronthaul SINR from the $l$th RRH under the MR and ZF schemes can be computed, respectively, as
\begin{align}
\operatorname{SINR}_l^{\, \operatorname{MR}}=\frac{\beta_l\Vert\mathbf{b}_l\Vert^4 p_{H,l}}{\sigma^2\Vert\mathbf{b}_l\Vert^2+\sum_{l^{\prime}\neq l}\beta_{l^{\prime}}\vert \mathbf{b}_l^{\prime} \mathbf{b}_l\vert^2 p_{H,l^{\prime}}},
\end{align}
and 
\begin{align}
\operatorname{SINR}_l^{\,\operatorname{ZF}}&=\frac{\beta_l p_{H,l}}{[(\mathbf{B}^{\dagger}\mathbf{B})^{-1}]_{l,l}}.
\end{align}
Based on the decoded signal $y_{C,l}$, we get the corresponding compression index $T_l$. Given $T_l$, the CU can determine the quantized signal $\widehat{\mathbf{y}}_{H,l}$. With point-to-point compression at the RRHs, in order for the CU to recover $\widehat{\mathbf{y}}_{H,l}$, according to the rate-distortion theory, the condition
\begin{align}\label{fronthaulCondition}
    \gamma_l(\mathbf{V}_k, \mathbf{\Omega}_l) &\triangleq \operatorname{log\, det} \left(\frac{\sum_{k=1}^{K} \mathbf{H}_{lk}\mathbf{V}_k \mathbf{V}_k^{\dagger}\mathbf{H}_{lk}^{\dagger} +\mathbf{I}_{N_{H,l}}+ \mathbf{\Omega}_l}{\mathbf{\Omega}_l}  \right)  \nonumber \\
    &= \operatorname{log\, det} \left(\sum_{k=1}^{K} \mathbf{H}_{lk}\mathbf{V}_k \mathbf{V}_k^{\dagger}\mathbf{H}_{lk}^{\dagger} +\mathbf{I}_{N_{H,l}}+ \mathbf{\Omega}_l\right) \nonumber \\
    &- \operatorname{log\, det} \left(\mathbf{\Omega}_l\right)
    \leq C_l
\end{align}
must be satisfied, where the left-hand side of the inequality is obtained by evaluating the mutual information of the test channel in (\ref{testChannelData}).
\subsection{Achievable Rate}
We now determine the achievable rate for each UE. First, define $\widehat{\mathbf{y}}_{H}\!\triangleq\!\left[\widehat{\mathbf{y}}_{H,1}^T, \dots, \widehat{\mathbf{y}}_{H,L}^T\right]^T$, $\mathbf{H}_k\!\triangleq\!\left[\mathbf{H}_{1k}^T, \dots, \mathbf{H}_{Lk}^T\right]^T$, $\mathbf{z}_{H}\!\triangleq\!\left[\mathbf{z}_{H,1}^T, \dots, \mathbf{z}_{H,L}^T\right]^T$ and $\mathbf{q}\!\triangleq\!\left[\mathbf{q}_{1}^T, \dots, \mathbf{q}_{L}^T\right]^T\sim\mathcal{CN}(0, \mathbf{\Omega})$ with $\mathbf{\Omega}\!=\!\operatorname{diag}(\{\mathbf{\Omega}_{l}\}_{l\in\mathcal{L}})$. Accordingly, we can write the received signal after quantization from all RRHs by plugging (\ref{accesLink}) into (\ref{testChannelData}) such that we obtain 
\begin{align}\label{ReceivedRRHafterq}
\widehat{\mathbf{y}}_{H}&=\sum_{k=1}^{K} \mathbf{H}_{k}\mathbf{V}_k \mathbf{s}_k+\mathbf{z}_{H}+\mathbf{q} \nonumber \\
&= \mathbf{H}_{k}\mathbf{V}_k \mathbf{s}_k+\sum_{j\neq k} \mathbf{H}_{j}\mathbf{V}_j \mathbf{s}_j+\mathbf{z}_{H}+\mathbf{q}.
\end{align}
By treating the interference as noise, the rate achievable by the $k$-th UE is given by
\begin{align}
R_k&\leq I(\mathbf{S}_k;\widehat{\mathbf{Y}}_{H}) \\
&=\operatorname{log\, det} \left(\mathbf{I}_{N_{H}}+ \frac{\mathbf{H}_{k}\mathbf{V}_k \mathbf{V}_k^{\dagger} \mathbf{H}_{k}^{\dagger}}{\sum_{j\neq k}\mathbf{H}_{j}\mathbf{V}_j \mathbf{V}_j^{\dagger} \mathbf{H}_{j}^{\dagger}+\mathbf{I}_{N_{H}}+\mathbf{\Omega} } \right) \nonumber \\
&= \operatorname{log\, det} \left(\sum_{k}\mathbf{H}_{k}\mathbf{V}_k \mathbf{V}_k^{\dagger} \mathbf{H}_{k}^{\dagger}+\mathbf{I}_{N_{H}}+\mathbf{\Omega}\right) \nonumber\\
&-  \operatorname{log\, det} \left(\sum_{j\neq k}\mathbf{H}_{j}\mathbf{V}_j \mathbf{V}_j^{\dagger} \mathbf{H}_{j}^{\dagger}+\mathbf{I}_{N_{H}}+\mathbf{\Omega}
\right) \\
&\triangleq f_k(\mathbf{V}_k, \mathbf{\Omega}_l). \label{userAchievableRate}
\end{align}


%

\subsection{Joint Precoding and Compression Optimization}
Having determined the achievable rate, we now aim at maximizing the weighted sum rate, with a weighting coefficient $w_k$, by optimizing the quantization noise covariance matrices $\mathbf{\Omega}_l$, the preocding matrices $\mathbf{V}_k$ and the transmit power allocation $p_{H,l}$ at each RRH. The weighting coefficient $w_k$ can be selected to give a priority to a specific user. It is important to note that the rate in (\ref{userAchievableRate}) is achievable if conditions (\ref{fronthaulCondition}) and (\ref{beamformingCondition}) are satisfied. Accordingly, the optimization problem can then be formulated as follows
\begin{align}
&&\underset{\underset{\{\mathbf{V}_k\}_{k\in\mathcal{K}}}{ \{\mathbf{\Omega}_l, p_{H,l}\}_{l\in\mathcal{L}}}}{\operatorname{maximize}} &\quad \sum_{k=1}^{K} w_k  f_k(\mathbf{V}_k, \mathbf{\Omega}_l)& \label{obj.function.pp}\\
&&\operatorname{subject\, to}&\quad \gamma_l(\mathbf{V}_k, \mathbf{\Omega}_l) \leq C_l,\quad &\forall l\in \mathcal{L} \label{compression.constraint}\\
&&&\quad \alpha_l( p_{H,l})\geq C_l,\quad &\forall l\in \mathcal{L} \label{fronthaul.constraint}\\
&&&\quad 0\leq p_{H,l}\leq P_{H}^{\operatorname{ max}},\quad &\forall l\in \mathcal{L} \label{RRHPowerBudget}\\
&&&\quad \mathbf{\Omega}_l \succeq 0 ,\quad &\forall l\in \mathcal{L} \label{Cov.constraint} \\
&&&\quad \operatorname{tr}\left( \mathbf{V}_k \mathbf{V}_k^{\dagger}\right) \leq P_{UE},\quad &\forall k\in\mathcal{K}.\label{UserPowerConstraint}
\end{align}

\begin{table*}
\vspace{1em}
\centering
\begin{align} \label{apprObjectiveFunction}
\widehat{f}_k(\mathbf{F}_k^{(t)}, \mathbf{F}_k^{(t-1)}, \mathbf{\Omega}_l^{(t)},  \mathbf{\Omega}_l^{(t-1)}) &\triangleq
\operatorname{log\, det} \left(\sum_{k}\mathbf{H}_{k}\mathbf{F}_k^{(t)} \mathbf{H}_{k}^{\dagger}+\mathbf{I}_{N_{H}}+\operatorname{diag}( \{\mathbf{\Omega}_{l}^{(t)}\}) \right)\nonumber\\ 	
&-\xi\left(\sum_{j\neq k}\mathbf{H}_{j}\mathbf{F}_j^{(t)} \mathbf{H}_{j}^{\dagger}+\mathbf{I}_{N_{H}}+\operatorname{diag}( \{\mathbf{\Omega}_{l}^{(t)}\}),
\sum_{j\neq k}\mathbf{H}_{j}\mathbf{F}_j^{(t-1)}\mathbf{H}_{j}^{\dagger}+\mathbf{I}_{N_{H}}+\operatorname{diag}( \{\mathbf{\Omega}_{l}^{(t-1)}\})   \right)
\end{align}
\end{table*}

\begin{table*}
\centering
\begin{align} \label{apprCommpressionFunction}
\widehat{\gamma}_l(\mathbf{F}_k^{(t)}, \mathbf{F}_k^{(t-1)}, \mathbf{\Omega}_l^{(t)},  \mathbf{\Omega}_l^{(t-1)})&\triangleq 
\xi\left(\sum_{k=1}^{K} \mathbf{H}_{lk}\mathbf{F}_k^{(t)}\mathbf{H}_{lk}^{\dagger} +\!\mathbf{I}_{N_{H,l}}+\! \mathbf{\Omega}_l^{(t)},
\sum_{k=1}^{K} \mathbf{H}_{lk}\mathbf{F}_k^{(t-1)}\mathbf{H}_{lk}^{\dagger} +\!\mathbf{I}_{N_{H,l}}+\! \mathbf{\Omega}_l ^{(t-1)}\right)
\!-\operatorname{log\, det}\!\left(\mathbf{\Omega}_l^{(t)}\right)
\end{align}
\end{table*}

\begin{table*}
\centering
\begin{align}\label{apprFronthaulFunction}
\widehat{\alpha}_l(P_{H,l}^{(t)}, P_{H,l}^{(t-1)} )\triangleq
\operatorname{log}_2\!\left(\!\sigma^2\Vert\mathbf{u}_l\Vert^2\!+ \sum_{l}\vert\mathbf{u}_l^{\dagger}\mathbf{g}_l\vert^2P_{H,l}^{(t)}\right)
\!-\!\xi\!\left(\!\sigma^2\Vert\mathbf{u}_l\Vert^2\!+\!\sum_{l^{\prime}\neq l}\vert\mathbf{u}_{l^{\prime}}^{\dagger}\mathbf{g}_{l^{\prime}}\vert^2P_{H,l^{\prime}}^{(t)},\,\sigma^2\Vert\mathbf{u}_l\Vert^2\!+\!\sum_{l^{\prime}\neq l}\vert\mathbf{u}_{l^{\prime}}^{\dagger}\mathbf{g}_{l^{\prime}}\vert^2P_{H,l^{\prime}}^{(t-1)} \!\right)
\end{align}
\medskip
\hrule
\end{table*}

\begin{algorithm}\label{version1}
Initialization: Select feasible $\mathbf{\Omega}_l^{(0)}, p_{H,l}^{(0)}, \mathbf{F}_k^{(0)}\quad\forall\, l\in\mathcal{L}$ and $\forall\, k \in \mathcal{K}$, $t=0$\;
\While{convergence is not met}{ 
$t \leftarrow t+1$\;
Update $\mathbf{\Omega}_l^{(t)}, p_{H,l}^{(t)}$ and $\mathbf{F}_k^{(t)}$ by solving the following comvex optimization problem 
\begin{align*}
&\underset{\underset{\{\mathbf{F}_k^{(t)}\}_{k\in\mathcal{K}}}{ \{\mathbf{\Omega}_l^{(t)}, p_{H,l}^{(t)}\}_{l\in\mathcal{L}}}}{\operatorname{arg\,max}}\sum_{k=1}^{K} w_k  \widehat{f}_k(\mathbf{F}_k^{(t)}, \mathbf{F}_k^{(t-1)}, \mathbf{\Omega}_l^{(t)},  \mathbf{\Omega}_l^{(t-1)})& &
\end{align*}
\begin{align*}
&\operatorname{subject\, to} && \nonumber\\
&\widehat{\gamma}_l(\mathbf{F}_k^{(t)}, \mathbf{F}_k^{(t-1)}, \mathbf{\Omega}_l^{(t)},  \mathbf{\Omega}_l^{(t-1)}) \leq C_l, & \forall l\in \mathcal{L} & \\
&\widehat{\alpha}_l(p_{H,l}^{(t)}, p_{H,l}^{(t-1)})\geq C_l, & \forall l\in \mathcal{L} \\
& 0<p_{H,l}^{(t)}\leq P_{H}^{\operatorname{ max}}, & \forall l\in \mathcal{L} &\\
&\operatorname{tr}\left( \mathbf{F}_k^{(t)} \right) \leq P_{UE}, & \forall k\in\mathcal{K}&\\
&\mathbf{\Omega}_l^{(t)} \succeq 0, & \forall l\in \mathcal{L}&;\ 
\end{align*}
}
Optimal solution: $\mathbf{\Omega}_l^\star = \mathbf{\Omega}_l^{(t)}$, $p_{H,l}^\star = p_{H,l}^{(t)}$, $\mathbf{F}_k^\star = \mathbf{F}_k^{(t)}$\;
From  $\mathbf{F}_k^\star$, obtain the optimal
solution $\mathbf{V}_k^\star$ using the eigenvalue decomposition.
\caption{Majorization-Minimization Algorithm for Joint Optimization of the Fronthaul and Access Links}
\end{algorithm}

By inspecting the objective function and the fronthaul constraints (\ref{compression.constraint}) and (\ref{fronthaul.constraint}), we can identify that this problem belongs to the class of a nonconvex problem. Since it is difficult to solve, we reformulate the problem into a class of difference of convex (DC) problem and solve it using the MM method. Adopting the approach given in \cite{6588350}, the first step is to make a change of variable $\mathbf{F}_k\triangleq  \mathbf{V}_k \mathbf{V}_k^{\dagger}$ such that we have a function $ f_k(\mathbf{F}_k, \mathbf{\Omega}_l)$ in the objective function (\ref{obj.function.pp}) and $\gamma_l(\mathbf{F}_k, \mathbf{\Omega}_l)$ in the first constraint function (\ref{compression.constraint}). Following the MM method, we approximate those functions by a sequence of concave/convex functions and solve the problem iteratively. At each iteration $t$, the objective function is approximated by concave function (\ref{apprCommpressionFunction}) given at the top of the next page, whereas constraints (\ref{compression.constraint}) and (\ref{fronthaul.constraint}) are approximated by convex functions (\ref{apprCommpressionFunction}) and (\ref{apprFronthaulFunction}), respectively. Specifically, the approximation is done by linearizing the nonconvex/noncocave terms using a Taylor series $\xi(\mathbf{X}, \mathbf{X}_0)$ around the optimal point $\mathbf{X}_0$ obtained from the previous iteration, which is defined by
\begin{align}
\xi(\mathbf{X}, \mathbf{X}_0)\triangleq \operatorname{log \, det}(\mathbf{X_0} )+\frac{1}{\operatorname{ln}2} \operatorname{tr}(\mathbf{X}_0^{-1} (\mathbf{X}-\mathbf{X}_0)).
\end{align}
In Algorithm \ref{version1} we summarize the steps to iteratively solve the resulting convex optimization problem.

\section{Numerical Results}\label{numResults}
\begin{figure}
\centering
\includegraphics[width=\columnwidth]{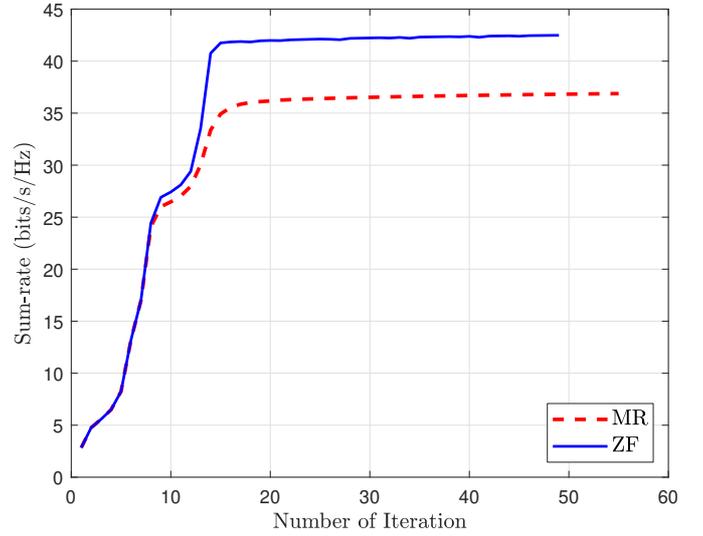}
\caption{The convergence of the proposed algorithm for MR and ZF fronthaul beamforming with $K=4, L=4, N_{U,k}=2, N_{H,l}=2$ and $N_C=200$.}
\label{ConMRZF}
\end{figure}

To evaluate the performance of our proposed system, we provide in this section some numerical simulations. We consider a square area of size 1 km$^2$, in which a CU is placed in the center. Around the CU, there are $L=4$ uniformly distributed RRHs and $K=10$ uniformly distributed UEs, each of which is equipped with $N_{U,k}=2$ and $N_{H,l}=3$ antennas, respectively. We make sure that the distance from RHHs to the CU is no closer than $10$ m, and the distance between RRHs is no closer than $100$ m. We consider the setup where the transmission between the fronthaul links and access links are separated in the time domain and operated at the same carrier frequency $1.9$ GHz with the system bandwidth $B=20$ MHz. As a result, we assume there is no self-interference between the transmit and receive antennas at the RRHs. For the fronthaul link, we consider LoS channel in (\ref{LoSgl}) with a free-space path loss model. For the access link, we consider Rayleigh fading channel with a rather realistic path loss model specified by 3GPP in \cite{3gppReleas16}, where we use the scenario of urban microcell (UMi) with non-LoS, and set the height of RRHs $h_{\operatorname{RHH}}=22.5$ m and the height of UEs $h_{\operatorname{UE}}= 1.5$ m. Throughout this section, we consider that all UEs have the same weighting coefficients, i.e., $w_k=1$, $\forall k\in \mathcal{K}$.

\begin{figure} 
\centering
\includegraphics[width=\columnwidth]{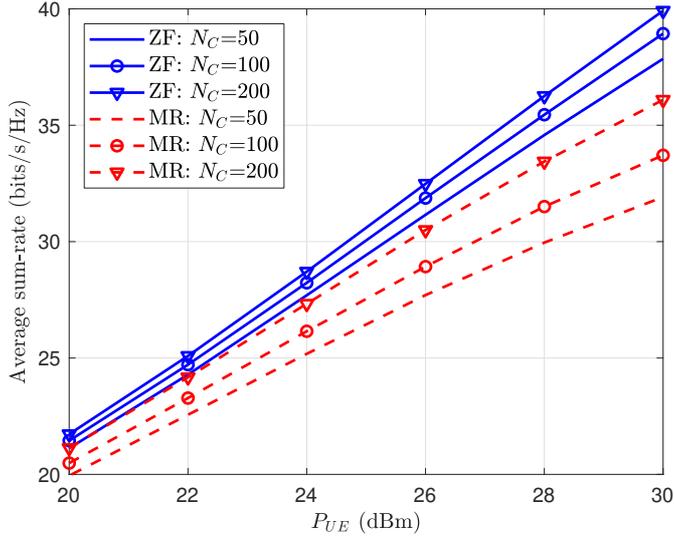}
\caption{The average sum-rate with respect to the UE transmit power $P_{UE}$ for for MR and ZF combining schemes, and for different number of antennas at the CU $N_C$ with $K=10, L=4, N_{U,k}=2, N_{H,l}=3$, and $P_{H}^{max}=30~\text{dBm}$. }
\label{SumRate_Pu}
\end{figure}

We first show in Fig. \ref{ConMRZF} the convergence of the proposed algorithm for the MR and ZF combining schemes. We can see that for both MR and ZF, the convergence is fast and as expected ZF outperforms MR. 

In Fig. \ref{SumRate_Pu}, we investigate the average sum-rate against the UEs transmit power $P_{UE}$, where the average sum-rate is computed using Algorithm \ref{version1} with $P_{H}^{\operatorname{ max}} = 30$ dBM over several channel realizations. Fig. \ref{SumRate_Pu} shows that the average sum-rate increases linearly with $P_{UE}$. As we increase the number of antennas at the CU $N_C$, the slope of the curve increases for both MR and ZF schemes. Moreover, we can observe that the ZF scheme generally outperforms the MR scheme. To achieve the same average sum-rate of 30 bits/s/Hz, the MR scheme requires 150 more CU antennas with slightly more UE transmit power than the ZF scheme. This is due to the well-known fact that ZF can suppress the interference between RRHs more effectively. 

\begin{figure} 
\centering
\includegraphics[width=\columnwidth]{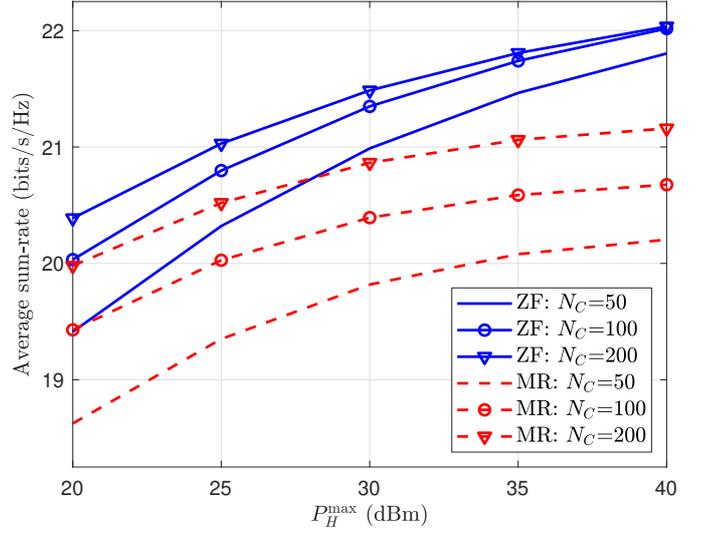}
\caption{The average sum-rate with respect to the maximum available transmit power at RRHs $P_{H}^{\operatorname{max}}$ for for MR and ZF schemes, and for different number of antennas at the CU $N_C$ with $K=10, L=4, N_{U,k}=2, N_{H,l}=3$, and  $P_{UE}=20 ~\text{dBm}$.}
\label{SumRate_PHmax}
\end{figure}

To further investigate the effect of the fronhaul links on the performance, we evaluate in Fig. \ref{SumRate_PHmax} the average sum-rate against $P_{H}^{\operatorname{ max}}$ for $P_{UE}=20 ~\text{dBm}$. As shown in the figure, the average sum-rate initially increases as $P_{H}^{\operatorname{ max}}$ increases, but then appears to be bounded at high $P_{H}^{\operatorname{ max}}$ for both  MR and ZF schemes. Note that we only specify the maximum available transmit power $P_{H}^{\operatorname{ max}}$, which is the same for all RRHs. However, different RRHs can have different transmit power $p_{H,l}$ as a variable of the optimization problem. Since Algorithm \ref{version1} should jointly give the optimal $p_{H,l}$ and $\mathbf{\Omega}_l$, increasing $P_{H}^{\operatorname{max}}$ is not expected to increase the sum-rate beyond a certain point to satisfy constraints (\ref{compression.constraint}) and (\ref{fronthaul.constraint}). This might explain the asymptotic behaviour observed in Fig. \ref{SumRate_PHmax}. In this regard, increasing the number of antennas $N_C$ and using a better fronthaul combining scheme can enlarge the feasibility set, as can be seen from the curve that shifts upwards. As we increase $N_C$, the gap between the curves appears to get smaller. It is shown more clearly for ZF, which indicates a bound in $N_C$. We leave the further investigation of this bound as our future work. 

\section{Conclusion}\label{conclusion}
This paper has studied the utilization of massive MIMO fronthaul in the uplink of the CRAN system. To maximize the sum-rate, we have jointly optimized the precoding matrices at the UEs, and the quantization covariance matrices and the transmit powers at the RRHs. An iterative algorithm to solve the non-convex optimization problem, based on the majorization minimization approach, has also been proposed. We have numerically investigated the performance of our proposed design for the MR and ZF schemes, where ZF has generally outperformed MR, as expected. Further, for both schemes, the sum-rate appears to show an asymptote with respect to the maximum available transmit power at the RRHs. These results give us an insight that the attempt to relax the fronthaul bottleneck by pushing the available power at RRH is less efficient. Rather, increasing the number of antennas at the CU brings more improvement.


%
%



\ifCLASSOPTIONcaptionsoff
  \newpage
\fi



\bibliographystyle{IEEEtran}
{\footnotesize \bibliography{References}}
\end{document}